\documentclass[11pt]{article}

\usepackage[a4paper,margin=1.2in]{geometry}

\usepackage{amsfonts,amsmath}
\usepackage{amsmath}
\usepackage{amssymb}
\usepackage{amsthm}
\usepackage{algorithm}
\usepackage{verbatim}
\usepackage{url}
\usepackage{enumitem}
\usepackage{algpseudocode} 
\usepackage{xcolor}
\usepackage{multirow}

\newcommand{\ket}[1]{\lvert #1 \rangle}

\newcommand{\abs}[1]{\lvert #1 \rvert}

\newcommand{\braket}[2]{\langle #1 \vert #2 \rangle}

\newcommand{\A}{\mathcal{A}}
\newcommand{\B}{\mathcal{B}}
\newcommand{\Q}{\mathcal{Q}}

\newcommand{\suppress}[1]{}

\newcommand{\Z}{{\mathbb Z}}
\newcommand{\F}{{\mathbb F}}

\begin{document}

\title{An exact quantum order finding algorithm 
\\
and its applications}

\author{
Muhammad Imran
\\
 Institute of Mathematics, Department of Algebra,
 \\
Budapest University of Technology and Economics,
\\
M\H{u}egyetem rkp. 3., Budapest, H-1111, Hungary.
\\
E-mail: \texttt{mimran@math.bme.hu
}}
\date{}

\maketitle
\begin{abstract}
    We present an efficient exact quantum algorithm for order finding problem when a multiple $m$ of the order $r$ is known. The algorithm consists of two main ingredients. The first ingredient is the exact quantum Fourier transform proposed by Mosca and Zalka in \cite{MosZal}. The second ingredient is an amplitude amplification version of Brassard and Hoyer in \cite{BraHoy} combined with some ideas from the exact discrete logarithm procedure by Mosca and Zalka in \cite{MosZal}. As applications, we show how the algorithm derandomizes the quantum algorithm for primality testing proposed by Donis-Vela and Garcia-Escartin in \cite{DoGa},  and serves as a subroutine of an efficient exact quantum algorithm for finding primitive elements in arbitrary finite fields. 

\end{abstract}

\section{Introduction}
Shor's quantum algorithm \cite{Shor} can determine the order of group elements efficiently, and it serves as the main tool for factoring integers. However, Shor's algorithm is polynomial-time in the expected sense, which means it may fail with a small probability and in the unlucky case may take a very long time to succeed, even may never terminate. The same case happens with Simon's algorithm \cite{Simon}. However, Brassard and Hoyer, in \cite{BraHoy}, came up with an exact quantum polynomial time for Simon's problem.  The Brassard-Hoyer algorithm utilizes a modified version of Grover's technique in \cite{Grover} to derandomize Simon's algorithm. Specifically, they propose a method that, assuming that we can construct a superposition in which the total squared amplitude of the "desired" constituents (intuitively, the probability of success) is $\frac{1}{2}$, boosts this success probability to $1$.

The question about the existence of exact quantum replacements for bounded quantum error probabilistic algorithms is a natural question, as it is analogous to derandomizing probabilistic classical algorithms. Besides, some earliest quantum algorithms that demonstrate the power of quantum computers, such as Deutsch-Jozsa procedure \cite{DeutschJozsa} and Bernstein-Vazirani problem \cite{BerVaz}, are exact. It is a difficult open question whether Shor's factoring algorithm can be derandomized. In \cite{MosZal}, Mosca and Zalka successfully derandomize Shor's algorithm for discrete logarithm problem in a cyclic group of known order. All previous exact quantum algorithms are uniform, which means the circuits for the algorithms can be classically computed in time polynomial in the logarithm of the inputs, see \cite{NishOzaUnif} for the details of uniform quantum circuits. In \cite{MoscaMSRI}, Mosca proposed a slightly less general computational model in which it is allowed the gate $\B_r$ with parameter $r$ when $r$ emerges \textit{during} the computation. Using that model, Mosca successfully derandomized Shor's factoring algorithm. The square-free decomposition algorithm of Lie et al. \cite{LPDSsqfree} also works in this computational model.

Here we consider the question whether Shor's order finding algorithm can be derandomized in the uniform computational model assuming some knowledge. As knowing a multiple of the order $\Z_m^*$ would factor $m$ in randomized classical polynomial time, finding orders of group elements with a known multiple of the order is not necessarily as hard as factoring, so a multiple of the order may be a good candidate for such a help. An important example where this help is available is the case of computing multiplicative orders (and testing primitivity) of elements of finite fields.  

\paragraph{The main results.} In this paper, we show that, given a multiple of the order, we can adjust the total squared amplitude of the "desired states" in Shor's order finding algorithm to $\frac{1}{2}$ and use the Brassard-Hoyer amplitude amplification to obtain an exact algorithm, see section \ref{xShor}. The idea of the exact version of Shor's order finding algorithm presented in this paper can be generalized to an exact quantum algorithm for \textit{hidden subgroup problem} in $\Z_m^n$, see \cite{ImrIva}. As applications, we show how the algorithm derandomizes the quantum algorithm for primality testing proposed by Donis-Vela and Garcia-Escartin in \cite{DoGa},  serves as a subroutine of an efficient exact quantum algorithm for finding primitive elements in arbitrary finite fields in a slightly less general computational model proposed by Mosca in \cite{MoscaMSRI}.

\section{Amplitude amplifications}

Both Mosca-Zalka amplitude amplification and Brassard-Hoyer amplitude amplification are based on the generalization of Grover search algorithm in \cite{BraHoyTap}. We present a brief review of the general amplitude amplification discussed by Brassad, Hoyer and Tapp in \cite{BraHoyTap}.

Given an algorithm $\A$ using no measurement, the amplitude amplification is a method to boost the success probability of the algorithm $\A$. On initial input $\ket{0}$, the algorithm $\A$ returns a pure superposition $\A \ket{0}=\sum_{i\in I}\ket{i}\ket{\Gamma_i}$ for some index set $I\subset \Z$. We consider $\chi:I \to \{0,1\}$ a Boolean function that separates the desired outcome states (all states $\ket{i}\ket{\Gamma_i}$ with $\chi(i)=1$) from the unwanted states (all states $\ket{i}\ket{\Gamma_i}$ with $\chi(i)=0$) as follows. Let $A=\{i\in I~|~\chi(i)=1\}$ and $B=\{i\in I~|~\chi(i)=0\}$. We write
$\A\ket{0}=\ket{\Gamma_a}+\ket{\Gamma_b}$, where
$$\ket{\Gamma_a}=\sum_{i\in A}\ket{i}\ket{\Gamma_i}\text{ and }\ket{\Gamma_b}=\sum_{i\in B}\ket{i}\ket{\Gamma_i}.$$
Hence the success probability of the algorithm $\A$ is $a=\braket{\Gamma_a}{\Gamma_a}=|\ket{\Gamma_a}|^2$. Therefore, the amplitude amplification operator for the algorithm $\A$ is defined as
\begin{equation}\label{eq1}
    \Q(\A, \chi, \phi, \varphi)=-\A S_0^{\phi}\A^{-1} S_{\chi}^{\varphi},
\end{equation}
where $S_{\chi}^{\varphi}$ and $S_{0}^{\phi}$ are phase changing operators defined by
$$S_{\chi}^{\varphi}\ket{i}\ket{\Gamma_i}=\left\{\begin{array}{ll}
\varphi\ket{i}\ket{\Gamma_i}
& \mbox{~if $\chi(i)=1$} \\
\ket{i}\ket{\Gamma_i} & \mbox{~otherwise,}
\end{array}\right.~~~~\text{and}~~~~S_{0}^{\phi}\ket{i}\ket{\Gamma_i}=\left\{\begin{array}{ll}
\phi\ket{i}\ket{\Gamma_i}
& \mbox{~iff $i=0$} \\
\ket{i}\ket{\Gamma_i} & \mbox{~otherwise,}
\end{array}\right.$$
with $\phi$ and $\varphi$ are complex number of unit length.

The operator $\Q$ is a generalization of Grover's iterations applied in his quantum search algorithm \cite{Grover}. Moreover, by setting $\phi=\varphi=-1$, we have for every $j\geq 0,$
$$\Q^j \A\ket{0}= k_j\ket{\Gamma_a}+l_j\ket{\Gamma_b}$$
where $$k_j=\frac{1}{\sqrt{a}}\sin ((2j+1)\theta) ~~~~~\text{and}~~~~ l_j=\frac{1}{\sqrt{1-a}}\cos((2j+1)\theta),$$
and $0 \leq \theta\leq \pi/2$ is defined so that $\sin^2\theta=a=|\ket{\Gamma_a}|^2$.

A natural question to ask whether it is possible to boost the success probability to certainty. It turns out there are positive answers to this question. In \cite{BraHoy}, Brassard and Hoyer present an optimal value for the parameters $\phi$ and $\varphi$, namely $\phi=\varphi=\sqrt{-1}$, such that whenever the success probability of an algorithm $\A$ is $\frac{1}{2}$, then one application of the amplitude amplification $\Q$ boosts the success probability to $1$. This is the approach that Brassard and Hoyer use to derandomize Simon's algorithm. Another positive answer is also presented in \cite{MosZal} by Mosca and Zalka. They use one application of $\Q$ with parameters $\phi=\varphi=-1$ to increase the success probability $\frac{1}{4}$ of an algorithm $\A$ to $1$. They use this variant of amplitude amplification to present an exact quantum Fourier transform and derandomize Shor's quantum algorithm for discrete logarithm. Therefore, one application of the exact quantum Fourier proposed by Mosca and Zalka requires three applications of the usual quantum Fourier transform.

As one may notice from some previous derandomizations, such as Simon's algorithm and Shor's discrete logarithm, the knowledge of the success probability of the algorithms makes the derandomizations possible. Therefore, in the next sections, we show how to use this amplitude amplification approaches to derandomize Shor’s order finding algorithm when a multiple of the order is known and to construct an exact algorithm for primitive finding problem.

\section{Exact quantum order finding algorithm}
\label{xShor}

The problem we consider is given an element group $x\in G$ and a multiple $m$ of the unknown order $r$ of $x$, determine the order $r$.  The first part of the algorithm is the standard Fourier sampling. We use here an exact version based on the exact quantum Fourier transform of Mosca and Zalka \cite{MosZal}. The standard Fourier sampling procedure maps $\ket{0}\ket{0}$ to $\sum_{k=0}^{m-1}\ket{k}\ket{\Gamma_k}$, where $\ket{\Gamma_k}=\frac{1}{m}\sum_{j=0}^{m-1}\omega^{kj}\ket{x^j}$ and $\omega=e^{2\pi i/m}$.
Write $j$ as $j_0+rj_1$ ($0\leq j_0\leq r-1$).
Then

$$\ket{\Gamma_k}=
\left\{\begin{array}{ll}
1/r\sum_{j_0=0}^{r-1}\omega^{kj_0}\ket{x^{j_0}}
& \mbox{~if $m/r\vert k$} \\
0 & \mbox{~otherwise,}
\end{array}\right.
$$ whence
$${\abs{\Gamma_k}}^2
=\left\{\begin{array}{ll}
1/r & \mbox{~if $m/r\vert k$} \\
0 & \mbox{~otherwise.}
\end{array}\right.
$$
In words, we have terms with $\ket{k}$ in the first register
only for those $k$ which are multiples of $m/r$.
Initially, any $k$ which is nonzero modulo $m$
is useful because $\frac{m}{\gcd(k,m)}$ is a proper divisor
of $r$. 
We have $\sum_{k\neq 0}\abs{\Gamma_k}^2=1-\frac{1}{r}$. However, fortunately, if we already know a divisor $d$ of $r$ then those values $k$ that give us new information are the non-multiples
of $\frac{m}{d}$. We have
$\sum_{kd\neq 0}\abs{\Gamma_k}^2=1-\frac{d}{r}$. The point is we do not know $r$.

The second part of the algorithm is based on the discussion in the previous paragraph. We maintain a divisor $d$ of $r$. We construct iterations of a procedure that increase $d$. Initially $d:=1$. As long as $d<r$, we find $k$ such that $dk\bmod{m}\neq 0$. Then we replace $d$ with $\frac{m}{gcd(m,k)}$ since this is another divisor of $r$ greater than $d$. Hence, $d$ keeps increasing as long as $d<r$ and it stops immediately when $d=r$ as $dk=0\bmod{m}$ for all $k$ if and only if $d$ is a multiple of $r$.

Assume $d<r$. Let $rep(dk)$ be the smallest positive integer representative of $dk\bmod{m}$. In this case, $rep(dk)=d\frac{m}{r}$ for all $k$. Then $rep(dk)$ divides $m$ and the $\frac{m}{rep(dk)}-1=\frac{r}{d}-1$ positive integers of the form $t  d\frac{m}{r}<m$ are the nonzero multiple of $d\frac{m}{r}$ modulo $m$. Note that if $\frac{r}{d}$ is even, then the integers of the form $t d\frac{m}{r}$ with $m/2\leq t  d\frac{m}{r}<m$ represent just half of multiples of $d\frac{m}{r}$ modulo $m$. However, if $\frac{r}{d}$ is odd, we need to add another multiple of $d\frac{m}{r}$ modulo $m$, say $d\frac{m}{r}$, with weight $\frac{1}{2}$. The problem is we do not know $d\frac{m}{r}$. However, fortunately, for at least one integer $0\leq j \leq \log_2m$, namely for $j=\lceil \log_2d\frac{m}{r}\rceil$, the interval $(0, 2^j]$ contains only $d\frac{m}{r}$ and no other multiple of $d\frac{m}{r}$ as if $j-1<\log_2d\frac{m}{r}\leq j$ then $d\frac{m}{r}\leq 2^j$ and $2d\frac{m}{r}>2^j$.

\begin{algorithm}[H]\label{alg1}
\caption{Exact quantum order finding algorithm}
\scriptsize
\begin{algorithmic}[1]
\State \textbf{Initialize:} $d \gets 1$, ${Found}\gets 1$;
\While{${Found > 0}$}
    \For{$j=-1,\ldots, \lfloor \log_2 m\rfloor$}\vspace{0.1cm}
	   \State{$\chi_j(k,b)=
		\left\{\begin{array}{ll}
			1 & \mbox{if $rep(dk)\geq\frac{m}{2}$ or
					$b=1$ and $0<rep(dk)\leq 2^j$;} \\
			0 & \mbox{otherwise;}
		\end{array}
		\right.$}
	\State{${\mathcal U}_j:
  		\ket{0}\ket{0}\ket{0}\ket{0}\mapsto 
\ket{\psi_j}=\frac{1}{\sqrt{2}}\sum
\ket{k}\ket{\Gamma_k}\ket{b}\ket{\chi_j(k,b)}$;\mbox{~~~~~~~~}}\Comment{where $k\in \{0,1,\dots, m-1\}$, $b\in \{0,1\}$.\mbox{~~~~~~}} 
	\State{Apply the amplitude amplified version of ${\cal U}_j$ to obtain$\ket{\psi'_j}=\sum
c'_{\chi_j(k,b)}\ket{k}\ket{\Gamma_k}\ket{b}\ket{\chi_j(k,b)}$.}
	\State{Look at the $\ket{k}$-register;}
	\If{$dk\neq 0\bmod{m}$} 
		\State{$d\gets \frac{m}{gcd(m,k)}$;}
    \Else
        \State{${Found}\gets {Found}-1$}
        \EndIf
    \EndFor
\EndWhile
\end{algorithmic}
\end{algorithm}
Each round consists of iterations for $j=-1,\dots,\lfloor \log_2m\rfloor$ instead of starting with index $j=0$ to cover both cases when $\frac{r}{d}$ is even and when $\frac{r}{d}$ is odd. The case when $\frac{r}{d}$ is even is covered at least once, when $j=-1$ where the interval $(0,2^j]$ does not contain any integer. While the case when $\frac{r}{d}$ is odd is covered at leat once, when $j=\lceil \log_2d\frac{m}{r}\rceil$.

As in each round before termination, the size of $d$ is increased by at least a factor $2$ and it stops immediately when $d=r$, we need at most $\lceil \log_2 r \rceil$ rounds of iterations. The overall number of calls to the exact Fourier transform or its inverse is $O(\log m\log r)$.

\section{Applications}\label{sec4}
\subsection{Exact quantum algorithm for primality testing}
There are various efficient primality testing algorithms for a given large integer $n$ based on the factorization of $n-1$, see for example \cite{pratt}. The algorithms are based on the Lucas-Lehmer condition in \cite{lehmerTest}. 
Donis-Vela and Garcia-Escartin in \cite{DoGa} present an alternative quantum algorithm for testing whether a given number $n$ is a prime or not, without factoring $n-1$.

The quantum algorithm proposed by Donis-Vela and Garcia-Escartin requires Shor's order finding algorithm. In fact, Shor's order finding algorithm is the only quantum part of the algorithm. The primality testing algorithm works as follows. To test primality of an integer $n$, we choose a random $1<x<n$. The condition $n$ to be composite is when $\gcd (x,n)>1$ or $x^{\frac{n-1}{2}}\not \equiv \pm 1 \bmod{n}$. If $x^{\frac{n-1}{2}}\equiv-1 \bmod{n}$, then we compute the order of $x$. If the order of $x$ is $n-1$, then $n$ is prime. Therefore, we can use the exact algorithm 1 to derandomize the quantum part of the primality testing algorithm because a multiple $n-1$ of the order of $x$ is known.

\subsection{Exact quantum algorithm for primitive finding problem}
It is well known that the density of primitive  elements in a finite field $\F_q$ is large enough so that the simple method of choosing a small number of elements in $\F_q$ at random is in fact a probabilistic polynomial time algorithm for finding primitive elements. However, the existence of a deterministic or even Las Vegas polynomial time algorithm for finding primitive elements in arbitrary finite fields is still an open question. Almost all known classical algorithms for testing a generator require factoring $q-1$. In this section, we present an exact quantum algorithm for finding primitive elements using the algorithm in section 2 as a subroutine.

Let $x$ be an element of $G=\F_q^*$. The exact algorithm consists of two parts. The first part is testing primitivity using the exact algorithm 1. The second part is an exact procedure to find an element in $G\setminus\langle x\rangle$.

The procedure for the second part is the following. We start with the uniform superposition $\ket{G}=\frac{1}{\sqrt{m}}\sum_{y\in G}\ket{y}$ where $m=q-1$. If $x$ is an element of order $r<m$, then
$$\Big|\frac{1}{\sqrt{m}}\sum_{y\in G\setminus \langle x\rangle}\ket{y}\Big|^2=1-r/m \geq 1/2,$$
and one can adjust the success probability to $\frac{1}{2}$ as follows.
Let $\B_r$ denote the unitary operator that rotates $\ket{0}$ to $\sqrt{1- \frac{m}{2(m-r)}}\ket{0}+\sqrt{\frac{m}{2(m-r)}}\ket{1}$. Therefore, by attaching an additional one qubit register to $\ket{G}=\frac{1}{\sqrt{m}}\sum_{y\in G}\ket{y}\ket{0}$ and applying $\B_r$ on the second register, we have the superposition
$$\frac{1}{\sqrt{m}}\sum_{y\in G}\ket{y}\Bigg(\sqrt{1- \frac{m}{2(m-r)}}\ket{0}+\sqrt{\frac{m}{2(m-r)}}\ket{1}\Bigg).$$
Let $\chi:G\times \{0,1\} \to \{0,1\}$ be the Boolean function defined as $\chi(y,b)=1$ if and only if $y\in G\setminus\langle x\rangle$ and $b=1$. Therefore, the total squared amplitudes of states $\ket{y}\ket{b}$ with $\chi(y,b)=1$ is $\frac{1}{2}$. Hence, we apply Brassard-Hoyer amplitude amplification to obtain a state $\ket{y}\ket{1}$ with probability $1$ where $y\in G\setminus \langle x\rangle$. Let $\cal P$ denote the exact procedure discussed previously.

Now we can find a primitive element in $\F_q^*$ as follows. Let $x\in \F_q^*$. For each round, we compute the order of $x$ using the exact algorithm 1. If the order of $x$ is less than $q-1$, then we do the following procedure: find an element $y\in \F_q^*\setminus \langle x\rangle$ and then replace $x$ with $\gcd(x,y)$.

\begin{algorithm}[H]
\caption{Algorithm for computing the probability of primitive elements}
\scriptsize
\begin{algorithmic}[1]
\State{Choose an elemment $x\in \F_q^*$};
\State{Compute the order $r$ of $x$ using \textbf{algorithm 1}};
\While{${r<q-1}$}
    \State{Use the procedure $\cal P$ to obtain $y\in \F_q^*\setminus \langle x\rangle$};
    \State{$x \gets{\gcd(x,y)}$};
\EndWhile
\end{algorithmic}
\end{algorithm}

As in each round, the size of $G$ is deacreased by at least a factor of $2$, we need at most $\log_2 q$ rounds to find a primitive element in $\F_q^*$. Therefore, the overall number of calls to the exact Fourier transform is $O(\log^3q)$.


\section{Discussions}
Shor's quantum algorithm for order finding in modulo $m$ requires $O((\log\log m)\log^3 m)$ quantum operations with $O(\log \log m)$ uses of modular exponentiations. The main subroutines of Shor's order finding algorithm are modular exponentiation and quantum Fourier transform. Modular exponentiation needs $O(\log m)$ multiplications, hence $O(\log^3 m)$ is the total complexity for a modular exponentiation while quantum Fourier transform is quadratic in $\log m$, see \cite{Shor}.

On the other hand, the exact algorithm presented in section 3 needs $O(\log^2 m)$ exact quantum Fourier transform of Mosca-Zalka \cite{MosZal} and the same complexity for modular exponentiations as in Shor's algorithm. As observed in section 2, one call to the exact quantum Fourier transform requires three calls to the standard quantum Fourier transform, then the number of the standard quantum Fourier transform used in the exact algorithm is still $O(\log^2 m)$. Thus, the exact algorithm requires $O(\log^4m)$ quantum operations. Therefore,  the overhead of the exact version from the standard Shor's algorithm is $O(\log m)$.

The exact algorithm in section \ref{xShor} does not imply the existence of an exact quantum factoring algorithm because the possible multiple $\varphi(n)$ of the order of $x$ is unknown in general. However, the algorithm can be used to derandomize several quantum algorithms such as primality testing and primitive finding algorithm where in the latter is the first deterministic algorithm for finding primitive element in any finite field $\F_p$.

\paragraph*{Acknowledgement.} The author is grateful to Gabor Ivanyos for his helpful comments and suggestions.
\bibliographystyle{alpha}
\bibliography{exshor}
\end{document}